# Dynamic second-order hyperpolarizabilities of Si$_2$C and Si$_3$C clusters using coupled cluster singles-and-doubles response approach

You-Zhao Lan*[1], Yun-Long Feng

*Zhejiang Key Laboratory for Reactive Chemistry on Solid Surfaces, Institute of Physical Chemistry, Zhejiang Normal University, Zhejiang, Jinhua 321004, China*

**Abstract**

We investigate the dynamic second-order hyperpolarizabilities $\gamma(-3\omega; \omega, \omega, \omega)$ (indicated by $\gamma^{THG}$) of the Si$_2$C and Si$_3$C clusters using the highly accurate coupled cluster singles-and-doubles (CCSD) response approach. The static $\gamma$ values of the Si$_2$C and Si$_3$C clusters are $1.99 \times 10^{-35}$ and $3.16 \times 10^{-35}$ esu, respectively. Similar to the α values, the $\gamma$ values of the Si$_2$C and Si$_3$C clusters are smaller than those of the Si$_3$ and Si$_4$ clusters, respectively, which is related to much smaller static $\gamma$ value of the C atom than the Si atom.

**1. Introduction**

The (hyper)polarizabilities of small semiconductor clusters, such as silicon (Si) and silicon carbide (SiC) clusters, have attracted much attention in the past 20 years [1–17]. Numerous researches have contributed to the isotropic dipole polarizabilities α of the Si clusters [1–10, 12, 16]. For small Si$_n$ ($n < 10$) clusters, the α values are greater than the bulk value and decrease with increasing the cluster size [1, 2, 9, 12, 16]; moreover the α values of these small clusters depend directly on the cluster size and indirectly on the HOMO–LUMO energy gap [8]. For mediate-size Si$_n$ ($n = 9 – 50$) clusters, experimental α values [12] vary strongly and irregularly with cluster size and fluctuate around the bulk value. For large Si$_n$ ($n = 60 – 120$), all experimental α values are smaller than the bulk value [12]. However, for the Si$_n$ ($n = 9 – 28$) clusters, Deng *et al.* [6] found that theoretical α values exhibit fairly irregular variations with the cluster size and all calculated values are higher than the bulk value. Similar theoretical results have been also obtained by Jackson *et al.* [7] for the Si$_n$ ($n = 1 – 21$) clusters. Therefore, for intermediate-size Si$_n$ ($n = 9 – 28$), some discrepancies exist between experimental and theoretical static α values and more investigations would be needed. Besides an interesting size-dependence of the α values, a shape-dependence has been also theoretically reported by Jackson *et al.* [4] for the Si$_n$ ($n = 20 – 28$) clusters. Our recent theoretical study showed that the α values of the SiC$_n$ and Si$_n$C ($n = 2 – 6$) clusters are larger than the bulk polarizability of 3C-SiC and interestingly lie between the dipole polarizabilities of the Si and C atoms [15].

For the first- and second-order hyperpolarizabilities (β and γ), a few studies have been performed on the Si$_n$ ($n = 3 – 8$ and 10) [Ref. 13 and 17] and Si$_n$C and SiC$_n$ ($n = 2 – 6$) [Ref. 15]. *ab initio* finite field calculations were carried out on the static γ values of small Si$_n$ ($n = 3 – 8$) clusters [17] and showed that clusters with even atom number increase the γ value in the size-dependence of the γ values, while ones with odd atom number decrease the γ value. The best theoretic static γ values of the Si$_3$ and Si$_4$ clusters are about $5 \times 10^{-35}$ esu based on highly accurate coupled cluster (CC) calculations [10, 14]. For the Si$_n$C and SiC$_n$ ($n = 2 – 6$) clusters, the C-rich clusters have lower *α* and larger *β* than the Si-rich clusters [15]. The size-dependence of the *β* values of the SiC$_n$ (n = 2 – 6) clusters, which have approximate Si-terminated linear chain geometry, is similar to that observed in π-conjugated organic molecules.

---
[1] Corresponding author: Youzhao Lan; Postal address: Zhejiang Key Laboratory for Reactive Chemistry on Solid Surfaces, Institute of Physical Chemistry, Zhejiang Normal University, Jinhua 321004, China; Fax: +086 579 82282269; E-mail address: lyzhao@zjnu.cn

The above mentioned experimental and theoretical studies show that investigating the (hyper)polarizabilities of semiconductor clusters is all along a topic of interest. Small SiC clusters exhibit similar *α* and *β* to the Si clusters. Little is known, however, about the *γ* of the SiC material. More recently, the coupled perturbed Hartree Fock (CPHF) calculations [18] for periodic system showed that the magnitude of $\chi_{iijj}^{(3)}$ (*i*, *j* = *x*, *y*, *z*) is of the order of $10^{-14}$ esu for cubic SiC bulk (experimental lattice parameter: 4.358 Å). In this paper, we investigate the static and dynamic *γ* of the Si$_2$C and Si$_3$C clusters. A highly accurate calculation method, the response theory within *ab initio* CCSD framework, is employed.

## 2. Theory and Computational Details

Most stable geometries of the Si$_2$C and Si$_3$C clusters were obtained from the literature [19] and reoptimized at the DFT/aug-cc-pVTZ level using the Gaussian 03 program [20] with the hybrid Becke3–Lee–Yang–Parr (B3LYP) functional. The vibrational frequencies were calculated to confirm that the final geometries are stable without an imaginary frequency. The final geometries and symmetries of the Si$_2$C and Si$_3$C clusters are shown in Fig. 1.

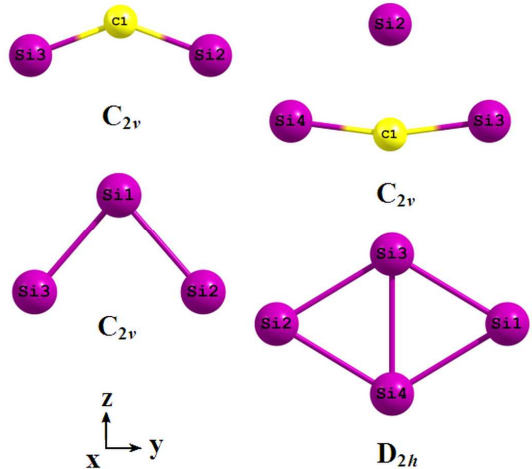

Figure 1. *Geometries of the Si$_2$C, Si$_3$C, Si$_3$, and Si$_4$ clusters.*

We used the CCSD response approach, as implemented in the Dalton 2.0 program [21] to calculate the *γ*. A detailed description of the response theory used to calculate the *γ* can be obtained from Ref. [22 and 23] and our previous work [14]. To obtain accurate *γ*, we mainly consider the choice of both coupled cluster models and basis sets. The inclusion of double excitation in the coupled cluster theory is proved to be important for the hyperpolarizability [22, 24]. In this study, we employ the CCSD response approach to calculate the dynamic hyperpolarizabilities *γ* for the third-harmonic generation (THG) technique. In the usual experiments, two components, $\gamma_\parallel$ and $\gamma_\perp$, are measured [25], where the optical field is polarized parallel and perpendicular to the static field, respectively. We focus on the scalar component of the tensor *γ*, which is defined by the isotropic average, $\gamma_\parallel = (1/15) \sum_{ij} (\gamma_{iijj} + \gamma_{ijij} + \gamma_{ijji})$, where *i*, *j* = *x*, *y*, *z*. In the static limit, $\gamma_\parallel$ is equal to the static isotropic average $\langle\gamma\rangle = (1/5) \sum_{ij} \gamma_{iijj}(0)$. For ease of choice of the basis sets, we provide the basis set dependence of $\gamma_{\parallel}(0)$ in Subsection 3.1 and select the d-aug-cc-pVDZ basis set, which has reached the basis set limit, for calculating the dynamic $\gamma$.

## 3. Results and Discussion

3.1. Basis set dependence of $\gamma_{\parallel}(0)$

Both the electron correlation effects and the quality of the basis set are important for the hyperpolarizability calculations [22, 24–27]. The coupled cluster theory can efficiently deal with the dynamical correlation effects. [22, 24, 28, 29] Here, we only perform the CCSD calculations because the CCSD model has been also proved to give accurate values for the optical properties based on the cubic response function, such as second-order hyperpolarizability and two-photon transition strength [22, 28, 29]. For the chose of the basis set, sufficient polarization and diffuse functions generally need to be involved in the basis set [14, 22, 26]. Both the Pople basis sets and the Dunning basis sets that include no diffuse functions (*e.g.*, 6-31G, 6-311G, 6-311G*, cc-pVDZ, and cc-pVTZ) would result in an irregularity in the sign or the value of $\gamma$ (Ref. [14]). To obtain accurate $\gamma$, at least one set of diffuse basis functions should be added in the basis set. Similar case occurs in the CCSD calculations of the two-photon absorption property [30] that is related to the cubic response function. Paterson *et al.* [28] showed that the unaugmented Dunning basis sets gives about one order of magnitude smaller two-photon transition strength ($\delta^{TPA}$) than the augmented ones. They also showed that the coupled cluster $\delta^{TPA}$ values based on the Pople basis sets are very poor and suggested that the Pople basis sets should be avoided in two-photon coupled cluster calculations. In our previous work [14], the Dunning basis set is also proved to be more suitable for the hyperpolarizability calculations than the Pople basis set.

Therefore, to estimate the basis set dependence of $\gamma$, we used the Dunning basis sets to calculate the $\gamma_{\parallel}(0)$ values of the $Si_2C$ and $Si_3C$ clusters. The calculated results are collected in Table 1. For comparison, we also provided the $\gamma_{\parallel}(0)$ obtained using the HF response calculations in Table 1. As expected, the unaugmented Dunning basis sets give erratic $\gamma_{\parallel}(0)$ values for both the CCSD and the HF response approaches. Addition of one set of diffuse functions significantly improves the results (see aug-cc-pVXZ: X = D and T). For the singly augmented basis sets, increasing the cardinality results in an increase of the $\gamma_{\parallel}(0)$ value (also see aug-cc-pVXZ: X = D and T). However, for both the doubly and the triply augmented basis sets, increasing the cardinality results in a slight decrease of the $\gamma_{\parallel}(0)$ value (see x-aug-cc-pVXZ: x = d and t, X = D and T). The $\gamma_{\parallel}(0)$ values based on the doubly augmented basis set appear to reach the basis set limit. In the CCSD calculations of $\delta^{TPA}$, Paterson *et al.* [28] showed that the $\delta^{TPA}$ values reach the basis set limit in the augmented triple-zeta basis sets and the singly and doubly augmented basis sets give very close $\delta^{TPA}$ values beyond the triple-zeta level. Further, for the HF response approach, which dose not deal with the electron correlation, the HF $\gamma_{\parallel}(0)$ values are smaller than the CCSD ones by an error of between 8% and 18%. On the basis of Paterson *et al.*'s work [28] and our present results, we select the d-aug-cc-pVDZ basis set for calculating the excited state properties and dynamic hyperpolarizabilities in order to reduce the computational costs and also owing to the non-availability of experimental results for the direct comparison for the $Si_2C$ and $Si_3C$ clusters.

Table 1. *Basis set dependence of χ(0) (×10$^{-35}$ esu) of the Si$_2$C and Si$_3$C clusters obtained using the CCSD and HF response calculations (1 au = 6.235 × 10$^{-65}$ C$^4$ m$^4$ J$^{-3}$ = 5.036 × 10$^{-40}$ esu)*

|  | CCSD | | HF | | Error[a] | |
|---|---|---|---|---|---|---|
| Basis | Si$_2$C | Si$_3$C | Si$_2$C | Si$_3$C | Si$_2$C | Si$_3$C |
| cc-pVDZ | –0.300 | –0.056 | –0.079 | 0.001 | | |
| cc-pVTZ | –0.060 | 0.270 | 0.074 | 0.274 | | |
| aug-cc-pVDZ | 1.607 | 2.674 | 1.316 | 2.212 | 0.18 | 0.17 |
| aug-cc-pVTZ | 1.788 | 2.802 | 1.558 | 2.549 | 0.13 | 0.09 |
| d-aug-cc-pVDZ | 1.992 | 3.157 | 1.636 | 2.670 | 0.18 | 0.15 |
| d-aug-cc-pVTZ | 1.955 | 2.952 | 1.724 | 2.707 | 0.12 | 0.08 |
| t-aug-cc-pVDZ | 2.014 | 3.160 | 1.661 | 2.675 | 0.18 | 0.15 |
| t-aug-cc-pVTZ | 1.961 | 2.963 | 1.730 | 2.719 | 0.12 | 0.08 |

[a] Error = (CCSD – HF)/CCSD

3.2. Dynamic second-order hyperpolarizabilities: $\gamma_{//}^{THG}$

Before the $\gamma_{//}^{THG}$, we investigate the absorption spectra of the Si$_2$C and Si$_3$C clusters. The absorption spectrum can provide important information on one- or multi-photon resonant absorption. One-photon or multi-photon absorption resonance enhancements will possibly lead to an irregularity in the sign and numerical value of the calculated hyperpolarizabilities and higher optical damage that should be avoided in nonlinear optical experiments [14, 27]. To obtain the CCSD absorption spectra, we calculated the excitation energies and oscillator strengths of the Si$_2$C and Si$_3$C clusters using the CCSD/d-aug-cc-pVDZ response calculations. Calculations were performed using the Dalton 2.0 program [21]. Then, we fitted the obtained excitation energies and oscillator strengths to the absorption spectra using Gabedit program [31] and a Lorentzian model with a half-bandwidth of 0.05 eV was used.

Figure 2 shows the absorption spectra of the Si$_2$C and Si$_3$C clusters, where the transition energy (TE) is less than 6.5 eV. For convenience of comparison, the calculated linear response density functional theory (LRDFT) absorption spectra [15] are also included in Fig. 2. The CCSD absorption spectra of the Si$_2$C and Si$_3$C clusters exhibit common characteristics of small semiconductor clusters with the number of atoms less than 10, such as Si$_n$ ($n$ = 3 – 10) [Ref. 14 and 16] and Ga$_n$As$_m$ ($n + m$ ≤ 10) [Ref. 32] clusters, that is, long absorption tails exist in the low-TE (<4.5 eV) region and strong allowed absorption peaks are located in the high-TE (>4.5 eV) region. In Fig. 2, the CCSD spectra are similar to the LRDFT spectra in terms of the peak positions and their envelopes except that the former exhibits a blue-shift (~0.1 eV). For clarity, Table 2 lists the transition energies (oscillator strengths) of the first excited state and the first strong allowed transition state for the Si$_2$C and Si$_3$C clusters. For the augmented Dunning basis sets (e.g. x-aug-cc-pVXZ: x = s, d, and t, X = D and T), the excitation energies based on different coupled cluster models (e.g. CCS, CC2, and CCSD) or different hybrid density functionals (e.g. B3LYP, BPW91, and B3P86) are very close [15, 28]. For instance, for the lowest excited state of the CH$_2$O molecule, Paterson *et al.* [28] showed that the exaction energies based on the CCSD response calculations with the aug-cc-pVDZ, d-aug-cc-pVDZ, and t-aug-cc-pVDZ basis sets are 4.006, 3.998, and 3.997 eV, respectively. Therefore, our present CCSD/d-aug-cc-pVDZ results can provide the reliable linear absorption properties.

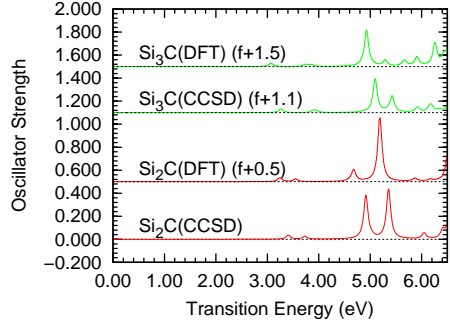

Figure 2. *Absorption spectra of the Si$_2$C and Si$_3$C clusters based on the CCSD/d-aug-cc-pVDZ and the B3LYP/aug-cc-pVTZ [Ref. 15] response calculations. Note that 0.5, 1.1, or 1.5 has been added to the oscillator strengths ( f ) to clearly display the plots.*

On the basis of the CCSD absorption spectra, we find that the Si$_2$C and Si$_3$C clusters have no obvious linear absorption in the region (< 4.5 eV) and have wide transparent region (< 3.0 eV). Therefore, we focus on the dynamic $\gamma_\parallel^{THG}$ of the Si$_2$C and Si$_3$C clusters with the applied field energies less than 2.0 eV. Figure 3 shows the dynamic $\gamma$ obtained using the CCSD/d-aug-cc-pVDZ response calculations. The three main components of the tensor $\gamma$ ($\gamma_{xxxx}$, $\gamma_{yyyy}$, and $\gamma_{zzzz}$) and the average $\gamma_\parallel^{THG}$ are provided, and the molecular orientations are shown in Fig. 1. Resonance enhancements at $\hbar\omega = 1.63$ eV (0.06 a.u.) are clearly observed for the Si$_2$C and Si$_3$C clusters. In the THG process, resonance enhancements will possibly occur at $\hbar\omega$, $2\hbar\omega$, and $3\hbar\omega$ applied field energies. A resonance results in a dispersion in the $\gamma$ values. For the Si$_2$C cluster, a resonance enhancement at 1.63 eV possibly is related to the $2\hbar\omega$ or the $3\hbar\omega$ resonance because the (1.63 eV × 2) and (1.63 eV × 3) are close to the transition energies of the first excited state (3.19 eV) and the first strong allowed transition state (4.92 eV) (Table 2), respectively. However, the oscillator strength of the first excited state is 0.0; thus the $2\hbar\omega$ resonance could be ignored. For the $3\hbar\omega$ resonance, we can see from Table 2 that there would be an actual absorption of photons because the oscillator strength ( f ) for the excited state of 4.92 eV is 0.3754. Similarly, for the Si$_3$C cluster, a resonance enhancement at 1.63 eV is related to the $3\hbar\omega$ resonance on the basis of the excited state of 5.10 eV (Table 2). The commonly employed laser wavelengths of 1064 (1.16) and 1907 nm (0.65 eV) are significantly different from the first strong resonance absorption energies at which the optical damage and thermal effects possibly occur; thus, the Si$_2$C and Si$_3$C clusters, as well as the Si$_3$ and Si$_4$ clusters [14], are potential candidates for THG nonlinear optical materials in the infrared range.

Table 2. *Transition energies (eV) and oscillator strengths (in parenthesis) of the first excited state and the first strong allowed transition for the Si$_2$C and Si$_3$C clusters.*

| Cluster | First excited state | | First strong allowed transition state | |
|---------|---------------------|---------------|---------------------------------------|---------------|
|         | CCSD                | LRDFT         | CCSD                                  | LRDFT         |
| Si$_2$C | 3.19 (0.0)          | 3.08 (0.0)    | 4.92 (0.3754)                         | 4.68 (0.0998) |
| Si$_3$C | 2.02 (0.0017)       | 1.87 (0.0015) | 5.10 (0.2912)                         | 4.93 (0.3112) |

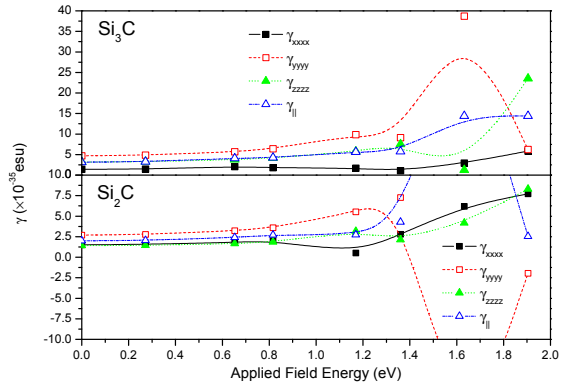

Figure 3. *Dynamic second-order hyperpolarizabilities γ of the $Si_2C$ and $Si_3C$ clusters based on the CCSD/d-aug-cc-pVDZ response calculations. (1 au = $6.235 \times 10^{-65}$ $C^4$ $m^4$ $J^{-3}$ = $5.036 \times 10^{-40}$ esu). For clarity, for the $Si_2C$ cluster, two data points ($γ_{//}$ and $γ_{yyyy}$ at 1.63 eV) are not shown, and they are $28.74 \times 10^{-35}$ and $-24.72 \times 10^{-35}$ esu, respectively.*

3.3. Comparison with the $γ_{//}^{THG}$ of the $Si_3$ and $Si_4$ clusters

As shown in Fig. 1, geometries of the $Si_3$ and $Si_4$ clusters are very similar to those of the $Si_2C$ and $Si_3C$ clusters. Researches on the static $α$ have shown that the static $α$ values of the $Si_3$ and $Si_4$ clusters are around 5.15 and 5.10 Å$^3$/atom based on different calculation methods [1, 2, 8, 9, 16], respectively, and that the static $α$ values of the $Si_2C$ and $Si_3C$ clusters are 4.04 and 4.00 Å$^3$/atom based on the LRDFT calculations [15], respectively. It has been shown that the size of the $α$ value is possibly related to the shape of the cluster or the energy difference between the molecular orbital levels. For example, for the $Si_n$ ($n$ = 3 – 10) clusters, Pouchan *et al.* [8] showed that more prolate structures are more polarizable and the $α$ value is directly related to the size of the energy gap between symmetry-compatible bonding and antibonding molecular orbitals. Jackson *et al.*[4] also found a clear shape-dependence for the calculated $α$ values of the $Si_n$ ($n$ = 20–80) clusters and clusters with prolate structures have systematically larger polarizabilities than those with compact structures. The reason why the static $α$ value of the $Si_2C$ cluster is smaller than that of the $Si_3$ cluster is much smaller static $α$ value of the C atom than the Si atom because the $Si_2C$ cluster has a similar geometry (or shape) to the $Si_3C$ cluster. Researches have shown that the C atom with a static $α$ value of 1.88 Å$^3$/atom [Ref. 33] is softer than the Si atom with that of ~5.54 Å$^3$/atom [Ref. 3]. Similar reason is for the $Si_3C$ and $Si_4$ cluster.

Motivated by the $α$, we make a comparison with the $γ_{//}^{THG}$ of the $Si_3$ and $Si_4$ clusters. A detailed study based on the CCSD response calculations for the $γ_{//}^{THG}$ values of the $Si_3$ and $Si_4$ clusters can be found in our previous work [14]. Finite field calculations for the static $γ_{//}$ of the $Si_3$ and $Si_4$ clusters can be found in the literature [10, 17]. Stable geometries of the $Si_3$ and $Si_4$ clusters are also shown in Fig. 1 based on the same optimization method as for the $Si_2C$ and $Si_3C$ clusters. In our previous work [14], the dynamic $γ_{//}^{THG}$ values are based on the CCSD/aug-cc-pVTZ response calculations. For a direct comparison, we recalculated the $γ_{//}^{THG}$ values of the $Si_3$ and $Si_4$ clusters using the CCSD/d-aug-cc-pVDZ response calculations. The obtained results for the dynamic $γ_{//}^{THG}$ values are provided for $ω$ = 0.0 (∞) and 0.65 eV (1907 nm) in Table 3. Further, nine main components of the tensor $γ$ (i.e., $γ_{iijj}$, $i, j$ = $x, y, z$) are provided along with the average $γ_{//}^{THG}$ values. For the $Si_3$ cluster, Champagne *et al.* [17] have used the finite field calculations of the RHF, MP2, MP3, MP4(DQ), MP4(SDQ), MP4(SDTQ), CCSD, and CCSD(T) with the 6-311+G* basis set to obtain the $\gamma_{//}(0)$ values of 2.48, 3.05, 2.76, 2.61, 2.72, 3.15, 2.74, and 3.03 ($\times 10^{-35}$) esu, respectively. In the case of the $Si_4$ cluster, they are 3.37, 4.52, 3.97, 4.00, 4.03, 4.42, 4.11, and 4.43 ($\times 10^{-35}$) esu, respectively. Moreover, Maroulis and Pouchan [10] obtained a $\gamma_{//}(0)$ value of $5.35 \times 10^{-35}$ esu by using the finite field calculations at the CCSD(T) level with a self-designed basis sets. Combining with our present results, we can find that the $Si_3$ and $Si_4$ clusters individually have a larger $\gamma_{//}(0)$ value than the $Si_2C$ and $Si_3C$ clusters. In the dynamic case, the $\gamma_{//}$ values at 0.65 eV in Table 3 exhibit the same size relationship as the static $\gamma_{//}$ values, i.e., $\gamma_{//}^{THG}(Si_3) > \gamma_{//}^{THG}(Si_2C)$ and $\gamma_{//}^{THG}(Si_4) > \gamma_{//}^{THG}(Si_3C)$. Actually, the dynamic $\gamma_{//}$ of the $Si_2C$, $Si_3C$, $Si_3$, and $Si_4$ clusters exhibit wide non-resonant optical region (Fig. 3 in this work and Fig. 2 in Ref. [14]) because their line absorption spectra have similar long absorption tails (<4.5 eV) (Fig. 2 in this work and Fig. 1 in Ref. [14]). In the non-resonant optical region, the size relationship for the static case remains valid for the dynamic $\gamma_{//}$ value that monotonically increases with an increase of the applied field energies (Fig. 3). Similar to the α, the $\gamma_{//}$ values of the $Si_2C$ and $Si_3C$ clusters are smaller $\gamma_{//}$ than those of the $Si_3$ and $Si_4$ clusters, respectively, because the C atom with a static $\gamma$ value of 2194.32 a.u. ($0.11 \times 10^{-35}$ esu) [Ref. 34] is much smaller than the Si atom with that of 430000 a.u. ($21.66 \times 10^{-35}$ esu) [Ref. 3].

Table 3. *Nine main components of the tensor γ and $\gamma_{//}^{THG}$ of the $Si_2C$, $Si_3C$, $Si_3$, and $Si_4$ clusters obtained using the CCSD/d-aug-cc-pVDZ response calculations. γ is expressed in $10^{-35}$ esu. 1 au = $6.235 \times 10^{-65}$ $C^4 m^4 J^{-3}$ = $5.036 \times 10^{-40}$ esu*

| Cluster | $\gamma_{xxxx}$ | $\gamma_{yyyy}$ | $\gamma_{zzzz}$ | $\gamma_{xxyy}$ | $\gamma_{xxzz}$ | $\gamma_{yyzz}$ | $\gamma_{yyxx}$ | $\gamma_{zzxx}$ | $\gamma_{zzyy}$ | $\gamma_{\parallel}^{THG}$ |
|---|---|---|---|---|---|---|---|---|---|---|
| $\hbar\omega = 0.0$ (eV) | | | | | | | | | | |
| $Si_2C$ | 1.51 | 2.69 | 1.41 | 0.89 | 0.49 | 0.80 | 0.89 | 0.49 | 0.80 | 1.99 |
| $Si_3C$ | 1.42 | 4.69 | 3.20 | 1.01 | 0.70 | 1.53 | 1.01 | 0.70 | 1.53 | 3.16 |
| $Si_3$ | 2.69 | 5.53 | 3.13 | 1.26 | 1.00 | 1.75 | 1.26 | 1.00 | 1.75 | 3.87 |
| $Si_4$ | 2.64 | 8.87 | 5.59 | 1.66 | 1.19 | 1.59 | 1.66 | 1.19 | 1.59 | 5.20 |
| $\hbar\omega = 0.65$ (eV) | | | | | | | | | | |
| $Si_2C$ | 1.80 | 3.23 | 1.66 | 1.10 | 0.58 | 0.99 | 0.89 | 0.57 | 0.95 | 2.40 |
| $Si_3C$ | 2.08 | 5.68 | 3.78 | 2.08 | 1.00 | 1.95 | 1.25 | 0.84 | 1.84 | 4.10 |
| $Si_3$ | 3.16 | 6.82 | 3.88 | 1.59 | 0.83 | 2.21 | 1.62 | 1.23 | 2.17 | 4.70 |
| $Si_4$ | 2.71 | 10.78 | 6.83 | 0.88 | 1.36 | 1.88 | 2.27 | 1.48 | 1.85 | 6.01 |

**4. Conclusions**

To our best knowledge, we have first investigated the dynamic $\gamma_{\parallel}^{THG}$ of the $Si_2C$ and $Si_3C$ clusters using highly accurate CCSD response approach. On the basis of Paterson *et al.*'s work [28] and our present results, we suggest that the reliable results should be obtained for the optical properties related to cubic response function using the CCSD with both the doubly and the triply augmented double- or triple-zeta Dunning basis sets response calculations. The dynamic $\gamma_{\parallel}^{THG}$ of the $Si_2C$ and $Si_3C$ clusters, as well as the $Si_3$ and $Si_4$ clusters [14], exhibit wide non-resonant optical region because there are long absorption tails in their linear absorption spectra. In the non-resonant optical region, similar to the α, the $\gamma_{//}$ value of the $Si_2C$ cluster is smaller than that of the $Si_3$ cluster, which is related to much smaller static $\gamma_{//}$ value of the C atom than the Si atom because these two clusters have a similar geometry (or shape). Similar reason is for smaller γ values of the Si$_3$C cluster than the Si$_4$ cluster. These clusters are expected to be potential candidates for third-order nonlinear optical materials in the infrared region. Our highly accurate theoretic results will be useful references for future experiments.

**Acknowledgements**

We appreciate the financial support from the Foundation of Zhejiang Key Laboratory for Reactive Chemistry on Solid Surfaces.